\documentclass[11pt,preprint]{article}

\usepackage{amsfonts}
\usepackage{amsmath}
\usepackage{amssymb}
\usepackage{graphicx}
\usepackage{cite}
\usepackage{hyperref}
\usepackage{geometry}

\usepackage{setspace}

\title{Nanofabrication beyond optical diffraction limit: Optical driven assembly enabled by superlubricity}
\author{\selectfont \small Liu Jiang-tao$^{1,2*\dag}$, Deli Peng$^{3,4*}$, Qin Yang$^1$, Ze Liu$^{4\ddagger}$, Zhenhua Wu$^{2,6\S}$\\  \scriptsize
$^{1}$College of Physics and Mechatronic Engineering,  Guizhou Minzu University, Guiyang 550025, China\\
\scriptsize  $^{2}$Key Laboratory of Microelectronic Devices and Integrated Technology, Institute of Microelectronics, \\ \scriptsize  Chinese Academy of Sciences, Beijing 100029, China\\
 \scriptsize $^{3}$
Institute of Superlubricity Technology, Research Institute of Tsinghua University in Shenzhen,  Shenzhen 518057, China\\
 \scriptsize $^{4}$
Department of Engineering Mechanics, Tsinghua University, Beijing 100084, China\\
 \scriptsize $^{5}$
 \scriptsize Department of Engineering Mechanics, School of Civil Engineering, Wuhan University,  Wuhan 430072, China\\
  \footnotesize $^{6}$
 \scriptsize School of Integrated Circuits, University of CAS, Beijing 100049, China\\
 \scriptsize   $^{*}$ Contributed equally to this work\\
 \scriptsize   $^{\dag}$Email: jtliu@semi.ac.cn
  \scriptsize  $^{\ddagger}$ Email: ze.liu@whu.edu.cn
    \scriptsize  $^{\S}$ Email: wuzhenhua@ime.ac.cn
 }

\begin{document}

\maketitle

\begin{abstract}
The optical manipulation of nanoparticles on superlubricity surfaces is investigated. The research revealed that, due to the near-zero static friction and extremely low dynamic friction at superlubricity interfaces, the maximum intensity for controlling the optical field can be less than 100 W/cm$^2$, which is nine orders of magnitude lower than controlling nanoparticles on traditional interfaces. The controlled nanoparticle radius can be as small as 5 nm, which is more than one order of magnitude smaller than nanoparticles controlled through traditional optical manipulation. Manipulation can be achieved in sub-microsecond to microsecond timescales. Furthermore, the manipulation takes place on solid surfaces and in non-liquid environments, with minimal impact from Brownian motion. By appropriately increasing dynamic friction, controlling light intensity, or reducing pressure, the effects of Brownian motion can be eliminated, allowing for the construction of microstructures with a size as small as 1/75 of the wavelength of light. This enables the control of super-resolution optical microstructures. The optical super-resolution manipulation of nanoparticles on superlubricity surfaces will find important applications in fields such as nanofabrication, photolithography, optical metasurface,  and biochemical analysis.
\end{abstract}




\section{Introduction}

The construction of solid surface structures finds important applications in various fields, including nanofabrication, semiconductor chips, micro-nanophotonics, catalysis, and biochemical analysis\cite{AOP7DB}. Among these, the use of optical methods, such as photolithography, for building specific surface structures offers advantages such as low cost, high efficiency, and suitability for mass production, making it the mainstream technology in practical manufacturing. Trillions of semiconductor chips are produced by constructing specific structures on silicon wafers using semiconductor photolithography processes. However, due to the wave nature of light, there is a diffraction limit when light passes through apertures or is focused, making it challenging to further reduce the size of the light spot\cite{https://doi.org/10.1002/pat.662,BornWolf:1999:Book}. As a result, the size of microstructures constructed using light is limited. To further reduce the size of these microstructures, it is necessary to operate with shorter-wavelength light. For advanced semiconductor processes, extreme ultraviolet (EUV) light is required for photolithography \cite{NP4WC,doi:10.1126/science.1071718,10.1116/1.2794048}. However, the production of EUV light sources and optical devices working in this wavelength range is extremely challenging, and their widespread adoption is limited.

In fact, while the size of a light spot is subject to diffraction limits, the intensity of light within the spot is not uniform. Typically, the center of the light spot has the highest intensity, and the intensity weakens exponentially as you move farther from the center. Researchers have harnessed this property of light spots and the fact that the force exerted by light on micro- and nanoparticles is proportional to the gradient of light intensity. This has allowed them to create optical tweezers\cite{MN5Neuman2008,NP5Juan2011,LSA10Zhang2021,LJ4OEA}, which capture and manipulate tiny particles using light, surpassing the diffraction limit of light. Optical tweezers have found widespread applications in fields such as biology\cite{doi:10.1126/science.1100603,FAVREBULLE2022932,Meijering2022}, medicine or chemistry\cite{Sullivan2020,JamaliNazariGhaffariVeluMoradi+2021+2915+2928}, and controlling surface morphology in materials\cite{LSA10Zhang2021,NL21Jiang2021,Ghosh2021}. However, due to the relatively weak forces generated by optical tweezers, high-intensity light is required to manipulate micro- and nanoparticles on traditional interfaces and construct surface microstructures\cite{Shakhov2015}. This limitation restricts the manipulation of micro- and nanoparticles.

In this regard, we have conducted research on the optical manipulation of nanoscale particles on superlubricity surfaces. Superlubricity refers to the phenomenon where friction between objects in relative motion is nearly zero or completely absent\cite{N563OH,LUO2021106092,AFM29ZW,PRL110YJ,NC8SWL,NC7EC,C161JL,N30HJ,MPeyrard_1983,ACSN17Bai2023,2308.02818v1}. Specifically, when two crystalline surfaces are in non-integral contact, structural superlubricity occurs, and the static friction is reduced to zero. Super lubrication can be achieved between various different media interfaces, such as graphite, molybdenum disulfide (MoS$_2$), gold, and hexagonal boron nitride (HBN)\cite{doi:10.1126/science.aad3569,NM21Liao2022, F11Ge2023,PhysRevLett.92.126101,F11Gao2023b,NR11Buch2018,F9Shi2021,AMI12Liu2020,BUZIO2021875}. Superlubricity can significantly reduce energy loss and wear between interfaces, enhancing the response speed of devices\cite{https://doi.org/10.1002/adfm.201806395,SC56Li2013}. Superlubricity has been employed in various fields, including high-speed self-recovering graphite blocks\cite{PRL110YJ}, high-speed vibrator\cite{PRL88ZQ,doi:10.1073/pnas.1922681117}, nanogenerators\cite{HUANG2020104494,NC12Huang2021}, high-speed switches\cite{CM2Wu2021}, hard disk drive\cite{10.1115/ISPS2016-9523}, and ultra-high-speed optical modulators\cite{ZHOU2021228,doi:10.1515/nanoph-2022-0185}.

Our research has revealed that, due to the absence of static friction on superlubricity surfaces, nanometer-sized particles can move freely on these surfaces, significantly reducing the difficulty of manipulating nanometer-sized particles within optical fields. Even in weak light fields, nanoparticles with a radius as small as 5 nm can be manipulated to converge at the center of the light spot under the influence of light gradient forces, thus overcoming the limitations imposed by the diffraction limit of light in traditional photolithography techniques for microfabrication. Moreover, owing to the minimal dynamic friction on superlubricity surfaces, nanoparticles exhibit extremely rapid response times and can be manipulated in the timescale of microseconds. Furthermore, as the contact occurs between solid-solid interfaces, the impact of Brownian motion on nanoparticles is minimal. By appropriately increasing the light intensity or the coefficient of dynamic friction, the positional deviations caused by Brownian motion can be reduced to as low as 1 nm at standard temperature and pressure.

\section{Theoretical model and calculation method}
\begin{figure}[t]
\begin{center}
\includegraphics[width=0.8\columnwidth]{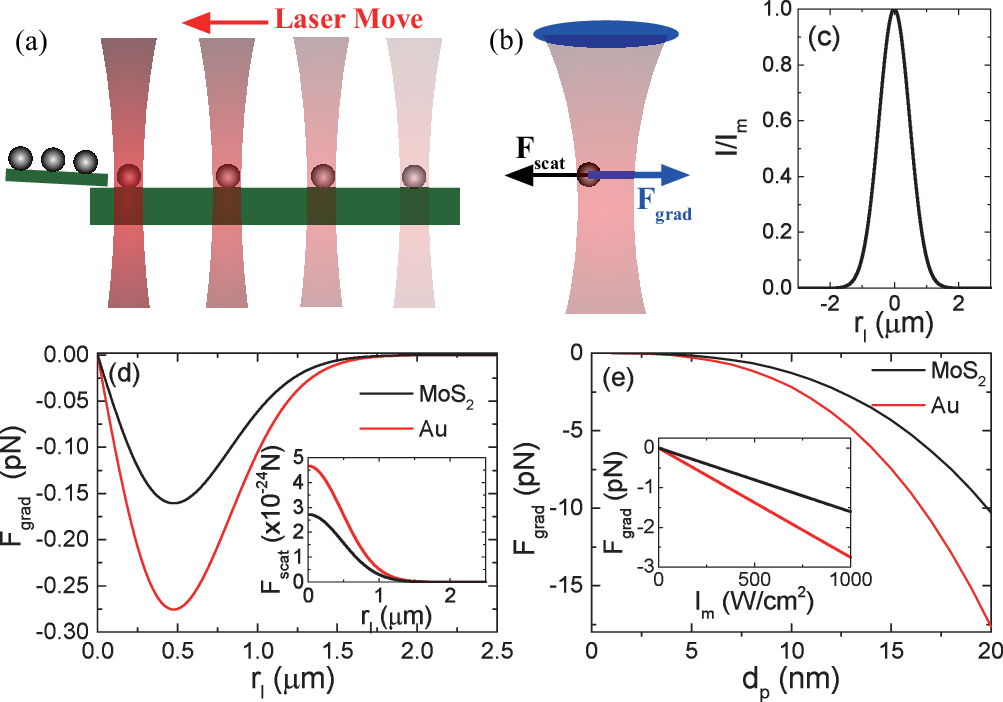}
\end{center}
\caption{(a) Schematic illustration of direct nanoparticle manipulation on a superlubricity surface using focused laser light. (b) Schematic representation of the forces acting on the nanoparticle in a Gaussian laser beam. (c) Axial distribution of light intensity at the waist of the Gaussian beam. (d) Variation of the light field gradient force $F_{grad}$ with axial distance from the nanoparticle, with an inset showing the variation of the light field scattering force $F_{scat}$ with axial distance. (e) Variation of the light field gradient force $F_{grad}$ with nanoparticle radius, with an inset illustrating the variation of $F_{grad}$ with the maximum light intensity $I_m$.}
\label{fig1}
\end{figure}

We initially investigated the laser direct writing mode, which involves focusing a laser beam to directly manipulate nanoscale particles on superlubricity surfaces. By moving different particles to specific positions, corresponding microstructures can be constructed, as shown in Fig. 1(a). Within the optical field, nanoscale particles are primarily subject to the action of two forces. Due to refraction of light and the electric charge polarization effect of the light field on the small spheres, a potential well is formed, pulling the small spheres toward the point of maximum light intensity and trapping them. The magnitude of this force is proportional to the gradient of light intensity and is referred to as the gradient force. When the particle size is much smaller than the wavelength of light, the magnitude of the gradient force can be expressed as follows\cite{HARADA1996529}
\begin{align}
F_{grad}=\frac{n^{2}_{s}r_{np}^{3}}{2}\left(\frac{\beta^{2}-1}{\beta^{2}+2}\right)^{2}\nabla I_{l},
\end{align}
where $n_{s}$ represents the refractive index of the medium surrounding the particle, $r_{np}$ is the particle radius, $\beta$ is the ratio of the particle refractive index to the refractive index of the surrounding medium, and $I_{l}$ is light intensity. The other force is primarily caused by scattering due to the absorption and re-emission of light by nanoscale particles\cite{HARADA1996529}
\begin{align}
F_{scat}=\frac{I_{l}n_{s}}{c}\frac{128\pi^{5}r_{np}^{6}}{3\lambda^{4}}\left(\frac{\beta^{2}-1}{\beta^{2}+2}\right)^{2},
\end{align}
where $\lambda$ represents the wavelength. The scattering force is proportional to the sixth power of the particle radius. Since this paper focuses on nanoscale particles with extremely small radii, the scattering force is negligible in comparison to the gradient force of light intensity. Lasers typically output as Gaussian beams and maintain their Gaussian beam profile even after focusing. The spatial distribution of the light field can be expressed as \cite{BornWolf:1999:Book}
\begin{align}
I(r_{l},z)=I_{m}\left[\frac{w_{0}}{w\left(z\right)}\right]^{2}exp\left[\frac{-2r_{l}^{2}}{w^{2}\left(z\right)}\right],
\end{align}
where $r_{l}$ represents the radial coordinate, $z$ is the axial coordinate with the coordinate origin at the beam waist, $I_m$ is the on-axis light intensity at the center, $w_0$ is the beam waist width, and $w(z)$ is the radius at which the light intensity drops to $1/e^2$ of the on-axis intensity. When the light wavelength is 750 nm and the beam waist radius is 0.95 $\mu m$, the axial distribution of light intensity at the beam waist exhibits a Gaussian profile. In this case, the light intensity distribution follows a Gaussian curve, with the highest intensity at the center and decreasing as you move further away from the axis.  In this way, the dynamic behavior of nanoparticles can be obtained by numerically solving Newton's laws of motion $\mathbf{F}=\mathbf{F}_{grad}+\mathbf{F}_{scat}+\mathbf{f}_{v}=m_{np}\mathbf{a}$, where $\mathbf{f}_{v}$ represents the frictional force.

\section{Results}
\subsection{Laser direct writing mode}
We initially investigated the laser direct writing mode, which involves focusing a laser beam to directly manipulate nanoscale particles on superlubricity surfaces. By moving different particles to specific positions, corresponding microstructures can be constructed, as shown in Fig. 1(a). Within the optical field,  nanoparticles are primarily influenced by two forces: the gradient force directing towards the region of highest optical intensity and the scattering force directing towards the region of lower optical intensity (Fig. 1(b)). Due to refraction of light and the electric charge polarization effect of the light field on the small spheres, a potential well is formed, pulling the small spheres toward the point of maximum light intensity and trapping them.  When the light wavelength is 750 nm and the beam waist radius is 0.95 $\mu m$, the axial distribution of light intensity at the beam waist exhibits a Gaussian profile (Fig. 1(c)). In this case, the light intensity distribution follows a Gaussian curve, with the highest intensity at the center and decreasing as you move further away from the axis.

Specifically, we are considering the optical control of MoS$_2$ nanoparticles on HBN surfaces. The HBN-MoS$_2$ lattice mismatch is relatively large ($\sim$24.6\%), resulting in extremely low friction coefficient, and there is no high-friction state that depends on layer stacking (independent of twist angles) due to interlayer isotropy \cite{NM21Liao2022}. Additionally, both HBN and MoS$_2$ exhibit very low light absorption in the near-infrared or partial red light spectrum. HBN has a low refractive index in this wavelength range ($\sim$2.1)\cite{pssb.201800417}, while MoS$_2$ has a higher refractive index (at a wavelength of 750nm, the refractive index is 5\cite{ARBeal_1979}), which is advantageous for enhancing light gradient forces. For the calculations provided below, unless otherwise specified, the wavelength is set to 750 nm, the nanoparticle radius $r_{np}$ is 5 nm, the beam waist radius $w_0$ is 0.95 nm, and the on-axis light intensity $I_{m}$ is set at 100 W/cm$^2$.

The force experienced by the 5 nm radius MoS$_2$ particles with $I_{m}=100$ W/cm$^2$ is shown in Fig. 1(d). The gradient force $F_{grad}$ reaches its maximum near the center of the beam waist, approximately 0.16 pN, with the force direction pointing along the axis. Due to the extremely small particle radius, the scattering force $F_{scat}$ , which is proportional to the sixth power of the particle radius, is on the order of $10^{-24}$ N, making it 11 orders of magnitude smaller than the gradient force $F_{grad}$. Therefore, in subsequent calculations, $F_{scat}$ can be neglected. For easier electrode formation and the reported occurrence of superlubricity between metal particles and some layered materials\cite{doi:10.1126/science.aad3569}, we also calculated the forces acting on nanoparticles of the same size, but made of gold (Au). Due to the presence of surface plasmon resonance in gold nanoparticles\cite{Wang2019SCIS,https://doi.org/10.1002/adom.201901631,Wen2022ACSP}, they experience a larger force. However, since the study focuses on controlling light with a longer wavelength, far from the plasmon resonance frequency, the gradient force $F_{grad}$ for gold nanoparticles, compared to MoS$_2$ particles, doesn't significantly increase in magnitude and is approximately 0.28 pN. Similarly, gold nanoparticles have minimal light absorption at a wavelength of 750 nm and their small size allows us to neglect the scattering force $F_{scat}$.

Next, we examined the force acting on the nanoparticles under equilibrium conditions (static) to understand how the optical field limits the nanoparticle's position. In a static scenario, with the interface placed horizontally, nanoparticles are primarily subjected to the forces of the light gradient and static friction. Only when the gradient force is greater than the static friction force can the nanoparticle's motion be ensured. In other words, the ratio of static friction force to the light gradient force determines the range of positions where the nanoparticle eventually comes to rest. For instance, when the static friction force is 10$^6$$G_m$ (where $G_m$ is the nanoparticle's weight, with a radius of 5 nm, $G_m$ equals $2.6\times10^{-20}$ N), the nanoparticle doesn't necessarily move to the center of the light spot. It may more likely come to rest outside $r_{l}=1.3$ $\mu m$ or near $r_{l}=40$ nm. Traditional interfaces typically have relatively high static friction forces, making it challenging for nanoparticles to move to the center of the light spot. This is also why traditional interfaces are less suitable for utilizing optical pressure to drive micro- and nanoparticles for achieving super-resolution structural control. When the static friction force is $10^5$$G_m$, its influence on the particle's position is minimal. In superlubricity interfaces where the static friction force is zero, nanoparticles will ultimately move to the center of the light spot regardless of how low the driving light intensity is.

\begin{figure}[t]
\begin{center}
\includegraphics[width=0.8\columnwidth]{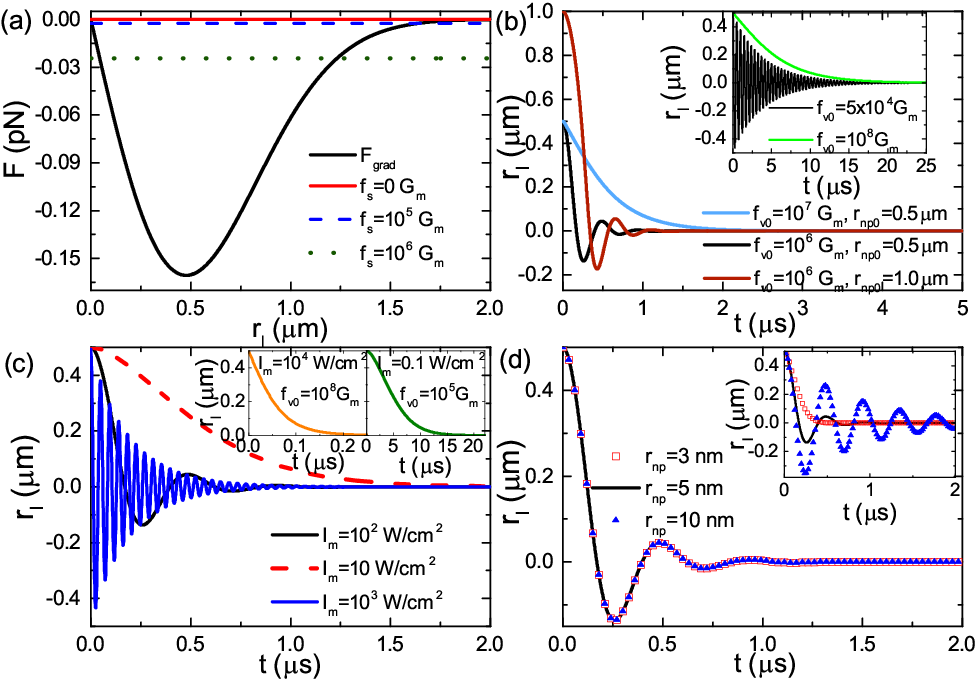}
\end{center}
\caption{(a) Comparison curve between gradient force and static friction force. (b) Nanoparticle's dynamic behavior for various initial positions and different dynamic friction forces. (c) Nanoparticle's dynamic behavior for different light intensity levels, with an inset illustrating the dynamic behavior for various light intensities and dynamic friction forces. (d) Nanoparticle's dynamic behavior for different particle radii, the inset depicts a scenario where dynamic friction is proportional to the particle's radius rather than its volume.}
\label{fig2}
\end{figure}

While on superlubricity interfaces, nanoparticles subjected to optical forces will eventually move to the center of the light spot, the magnitude of the dynamic friction force will affect the time it takes for the nanoparticles to reach the center. When the maximum light intensity $I_m$ is 100 W/cm$^2$, the dynamic behavior of nanoparticles under different initial positions and different dynamic friction forces is illustrated in Fig. 1(b).
 Dynamic friction increases with the increase in speed.  At relatively high speeds, the dynamic frictional force on the superlubricated interface is proportional to the velocity $v_{np}$ \cite{Guerra2010,PRL110YJ,PRB85LZ,2305.19740v1}, meaning $f_{v} = k_{f}v_{np}$. When $v_{np} = 1$ m/s, the friction force is represented as the friction constant $f_{v0}$. When the friction constant $f_{v0}$ is very low, for example, $f_{v0} = 5 \times10^4 G_m$, the optical driving force significantly exceeds the friction force. In this scenario, the kinetic energy gained by the nanoparticles from the optical force is not easily dissipated, resulting in multiple back-and-forth oscillations near the central point before stabilizing. When the friction constant $f_{v0}$ is relatively large, such as $f_{v0} = 10^8G_m$ (2.6 pN), the velocity of optically driven nanoparticles is lower, and it also takes a longer time to reach the equilibrium position.
When $f_{v0}$ is in the range of 10$^6 G_m$ to 10$^7 G_m$, the combined action of optical forces and friction dissipation allows nanoparticles to rapidly (within 1-2 $\mu s$) reach the center and stabilize. At this point, the average velocity of nanoparticles is roughly around 0.5 m/s. Furthermore, since the dynamic friction force is proportional to the velocity, the relationship between the final stabilization time and the initial position is not significant.

When $f_{v0} = 10^6 G_m$, the dynamic behavior of nanoparticles under different optical driving intensities is depicted in Fig. 1(c). The variation in light intensity alters the optical driving force. When the light intensity is weak (e.g., $I_{m} = 10$ W/cm$^2$), the driving force is smaller, and the nanoparticles have less kinetic energy, making them less prone to oscillations, but it takes longer for them to reach the center. Conversely, when the light intensity is strong (e.g., $I_{m}  = 10^3$ W/cm$^2$), the opposite occurs. Thus, under different dynamic friction forces, employing different optical driving intensities can avoid oscillations and enable rapid equilibrium. For instance, in the presence of higher dynamic friction ($f_{v0} = 10^8 G_m$), if a strong optical field ($I_{m}  = 10^4$ W/cm$^2$) is used, equilibrium can be reached in a very short time (0.1-0.2 µs), as shown in the left inset of Fig. 1(c). Appropriately increasing dynamic friction and the intensity of the driving light can enhance control speed. When dynamic friction ($f_{v0} = 10^6 G_m$) is low, reducing the light intensity ($I_{m}  = 0.1$ W/cm$^2$) can also prevent nanoparticle oscillations and lead to a faster center position, as shown in the right inset of Fig. 1(c). This way, reducing dynamic friction can significantly decrease the light intensity required for driving. Although the dynamic friction constant $f_{v0}$ varies considerably with temperature and specific production processes in practical manufacturing, the research results indicate that even with several orders of magnitude of change in $f_{v0}$, normal control of nanoparticles can be achieved by adjusting the light field intensity. This is advantageous for practical applications of optical nanoparticle manipulation.

As the particle radius increases, the driving force increases because the optical gradient force is proportional to the cube of the particle radius. However, the particle mass ($G_m$) is also proportional to the cube of the particle radius. When the dynamic friction force is normalized by the particle mass $G_m$ and remains constant, the driving force, dynamic friction force, and particle mass all scale proportionally with the cube of the particle radius. As a result, the acceleration remains constant, and the dynamic behavior of the nanoparticles is consistent.
In practical situations, superlubricity dynamic friction is generally related to the boundary length and varies less with changes in particle radius \cite{PhysRevLett.111.235502}. In this case, when the particle radius increases, the gradient force increases significantly, while the frictional force increases only slightly. Similar to the previous situation where the gradient force was large and the frictional force was small, the particle will undergo multiple vibrations before reaching equilibrium (inset of Fig. 2(d)). However, as mentioned earlier, one can still adjust the light intensity appropriately to achieve normal control of nanoparticles in such cases.

\begin{figure}[t]
\begin{center}
\includegraphics[width=0.8\columnwidth]{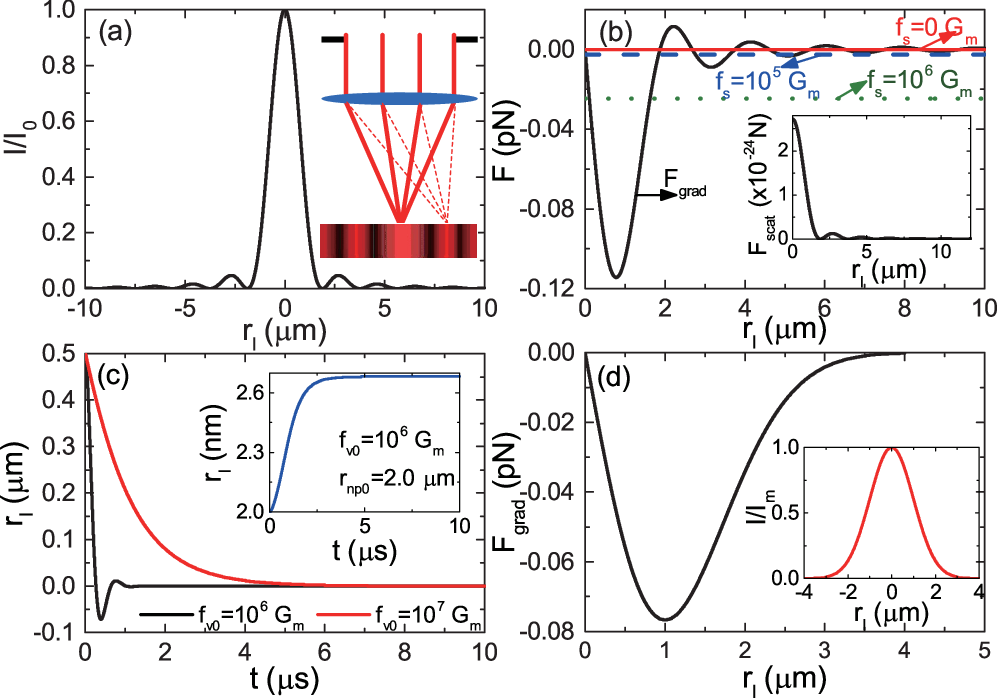}
\end{center}
\caption{(a) Intensity distribution curve of the slit diffraction pattern, with an inset illustrating the Fraunhofer diffraction. (b) Comparison curve between optical gradient force and static friction force, with an inset for optical scattering force. (c) Nanoparticle's dynamic behavior for an initial position of 0.5 micrometers under different dynamic friction forces, with an inset showing the nanoparticle's dynamic behavior for an initial position of 2 micrometers. (d) Gradient force of the Gaussian slit's Fraunhofer diffraction optical field, with an inset depicting the intensity distribution curve.}
\label{fig3}
\end{figure}

\subsection{Photolithography mode - Fraunhofer diffraction}
In addition to controlling individual nanoparticles through laser direct writing, microscale projection methods are commonly employed in optical surface processing techniques such as photolithography. One of these methods is called proximity projection exposure. In proximity projection, patterns from a mask or photomask are miniaturized and then exposed onto a photoresist. This approach allows for the simultaneous creation of lines or patterns, increasing the speed of constructing surface structures. In this type of exposure, the primary limitation comes from diffraction effects, particularly those caused by features like slits, making it challenging to further reduce line widths.

In response to this, we have also examined the influence of diffracted beams from slits on the control of nanoparticles on superlubricity surfaces. The intensity distribution of light due to Fraunhofer diffraction\cite{BornWolf:1999:Book}, perpendicular to the direction of the slit, is illustrated in Fig. 3(a). Similar to a Gaussian beam, the intensity of Fraunhofer diffraction is strongest at the center position and rapidly decreases as you move farther from the center. What sets it apart is the presence of a series of secondary peaks in addition to the diffraction main peak. Furthermore, the intensity of Fraunhofer diffraction remains constant along the direction of the slit.

When the central field strength $I_{m}= 100$ W/cm$^2$, the force exerted on 5 nm radius MoS$_2$ particles is depicted in Fig. 3(b). The gradient force $F_{grad}$ reaches its maximum value near the center of the primary peak, approximately 0.11 pN, with the force pointing towards the axis. The reason for $F_{grad}$  being smaller than in a Gaussian beam is that the diffraction width is wider in this case, leading to a reduced gradient in the light field. The scattering force $F_{scat}$ is also much smaller than $F_{grad}$ , by approximately 11 orders of magnitude, and can be neglected. In the equilibrium state (static), the comparison between static friction force and the gradient force of the light field also determines the range of final nanoparticle stopping positions.
Unlike the Gaussian beam, due to the presence of secondary peaks, even when the static friction force is equal to zero, multiple equilibrium points (intersections) exist.

The dynamic behavior of nanoparticles under various initial positions and different dynamic friction forces is shown in Fig. 3(c). When $f_{v0} = 10^6 - 10^7 G_m$, nanoparticles initially positioned 0.5 $\mu m$ away from the center, under the combined action of the light field force and friction dissipation, will rapidly reach the center and stabilize. However, if the initial position is far from the center but close to the center of a secondary peak, such as $r_{np0} = 2$ $\mu m$, the nanoparticle will stabilize at the position with the highest intensity of the secondary peak (inset of Fig. 3(c)). To avoid this situation, it is possible to modify the patterns on the mask or photomask, for example, by using Gaussian slits, which would result in a Gaussian distribution of the Fraunhofer diffraction field (inset of Fig. 3(d)). Its gradient force distribution curve is consistent with that of a Gaussian beam. In this case, the specific mechanics and dynamics are similar to the laser direct writing mode, and are therefore not discussed in detail. In practical applications, it is also possible to design photomasks using computational lithography methods such as inverse lithography technology to reduce or eliminate the effects caused by secondary light diffraction\cite{OE19Shen:11}.

\begin{figure}[t]
\begin{center}
\includegraphics[width=0.8\columnwidth]{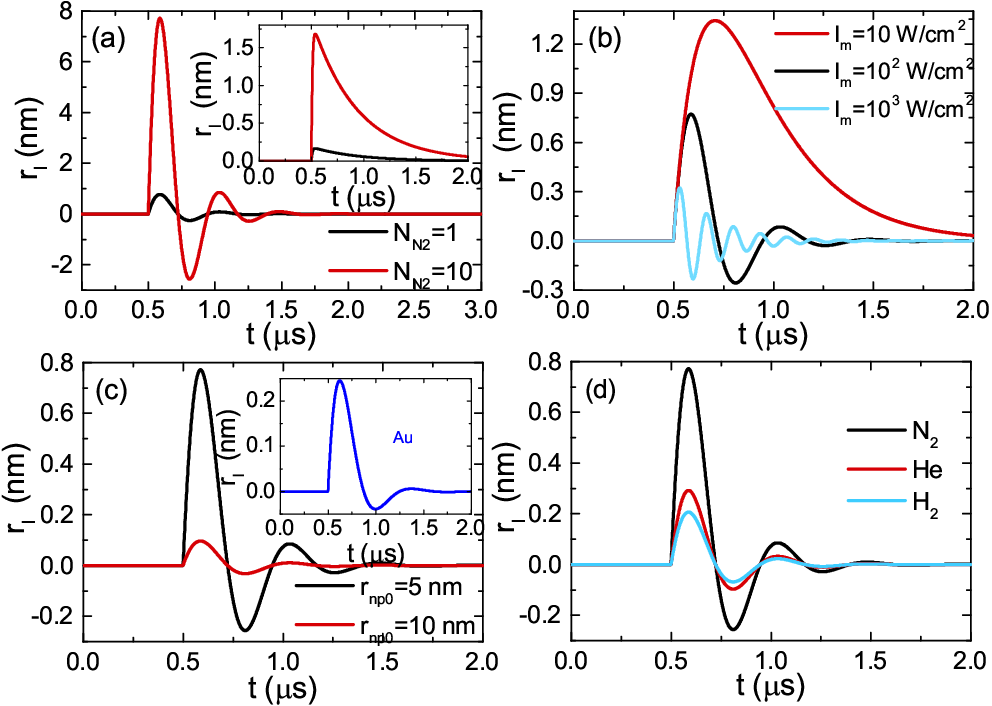}
\end{center}
\caption{(a) Nanoparticle's dynamics for different numbers of N$_2$ molecule collisions when the dynamic friction constant, $f_{v0}$, is $10^6 G_m$, with an inset for $f_{v0}$ equal to $10^7 G_m$. (b) Nanoparticle's dynamics under single collisions for various light intensities. (c) Nanoparticle's dynamics under single collisions for different particle radii, with an inset depicting the dynamics of Au nanoparticles. (d) Nanoparticle's dynamics under single collisions with different gas molecules.}
\label{fig4}
\end{figure}

\subsection{The effects of Brownian motion}
In optical trapping and other optical pressure control systems, one important factor that affects optical manipulation is Brownian motion caused by random molecular collisions in the environment\cite{doi:10.1126/science.1189403}. Traditional optical traps are typically performed in liquids, where the higher density of the liquid results in significant Brownian motion effects. Consequently, stronger optical intensities are usually required to trap micro- and nanoparticles. For superlubricity surfaces, the operations can be carried out in air, where the density is lower, and therefore, the impact of Brownian motion is reduced. We have also conducted research on this aspect.

To better understand the physical processes of Brownian motion, we initially studied the impact of single molecule collisions on nanoparticles. We first examined the dynamic behavior of 5 nm MoS$_2$ nanoparticles under collisions with nitrogen molecules (N$_2$) at room temperature. When a molecule collides with a nanoparticle, it almost rebounds at its original velocity while transferring a minimal amount of energy to the nanoparticle. After the collision, the nanoparticle's velocity is approximately $2M_{N_2}V_{N_2}/M_{np}$, where $M_{N_2}$ and $V_{N_2}$ are the mass and velocity of the N$_2$ molecule, and $M_{np}$ is the mass of the nanoparticle. For example, after a collision lasting 0.5 microseconds, the nanoparticle gains an instantaneous velocity and begins to move. However, under the influence of gradient forces and frictional forces, it eventually returns to its center position.

When the friction constant $f_{v0}=10^6 G_m$, a single molecule collision ($N_{N_2}=1$) results in a maximum displacement of approximately 0.77 nm (Fig. 4(a)). When the friction constant $f_{v0}$ is $10^7 G_m$, a single molecule collision leads to a maximum displacement of 0.17 nm (the inset of Fig. 4(a)). This is because friction hinders the nanoparticle's motion and dissipates the kinetic energy gained during the collision. When ten molecules collide simultaneously at the same speed, with a friction constant $f_{v0}$ of $10^6 G_m$ (or $10^7 G_m$), the maximum displacement of the nanoparticle is about 7.7 nm (or 1.7 nm). Due to the nanoparticle's rapid return to equilibrium within microseconds and its extremely small size, if the gas pressure is around 0.1 Pa (close to the lower limit of medium vacuum degree), there are approximately one collision. Therefore, under such gas pressure conditions, the deviations caused by molecular collisions can be neglected.

Increasing the intensity of the optical field can create a stronger gradient force field, which inhibits the nanoparticle from deviating from the center position. When the optical intensities are 10 W/cm$^2$ and 1000 W/cm$^2$, the maximum displacements of the nanoparticle are 1.3 nm and 0.31 nm, respectively (Fig. 4(b)). When the nanoparticle's radius increases, its mass increases, resulting in smaller velocities gained during collisions. Additionally, larger nanoparticle radii lead to stronger gradient forces on the nanoparticle. Both of these factors contribute to reducing the maximum displacement of the nanoparticle. When the nanoparticle's radius is 5 nm (10 nm), the maximum displacement is approximately 0.77 nm (0.096 nm), as depicted in Fig. 4(c). For Au nanoparticles, due to their larger mass and stronger gradient force resulting from their size, the maximum displacement for 5 nm nanoparticles is 0.24 nm (inset of Fig. 4(c)). Smaller molecular masses result in smaller momentum for the same kinetic energy, which leads to smaller velocities gained by the nanoparticle after collisions, resulting in smaller displacements. For environmental molecules N$_2$, He, and H$_2$, the maximum displacements of the nanoparticle are 0.77 nm, 0.29 nm, and 0.21 nm, respectively (Fig. 4(d)).

Building upon single-molecule collision simulations, we also simulated the impact of random molecular collisions, representing Brownian motion, on the nanoparticle's motion. Due to the extremely small nanoparticle radius and the low air density, we used very small time intervals in our calculations to ensure that the probability of a molecule-nanoparticle collision within a single time interval was less than 0.5, to avoid the influence of stochastic bias. For computational convenience, we only considered radial momentum transfer. The results are shown in Fig. 5, where we used a friction constant of $f_{v0} = 10^7 G_m$ and N$_2$ gas molecules in the calculations. Under the influence of collisions with environmental gas molecules, the nanoparticle exhibits irregular motion as it approaches the equilibrium position. However, due to the limiting effect of the gradient force field of the optical field, it doesn't deviate too far from the equilibrium position. When the air pressure is at 1 atmosphere, the nanoparticle's maximum displacement is close to 60 nm (Fig. 5(a)). When the air pressure is reduced, the probability of molecular collisions decreases, leading to a smaller maximum displacement. At air pressures of 0.1 (0.01) atmospheres, the maximum displacement is approximately 20 nm (5 nm).

\begin{figure}[t]
\begin{center}
\includegraphics[width=0.8\columnwidth]{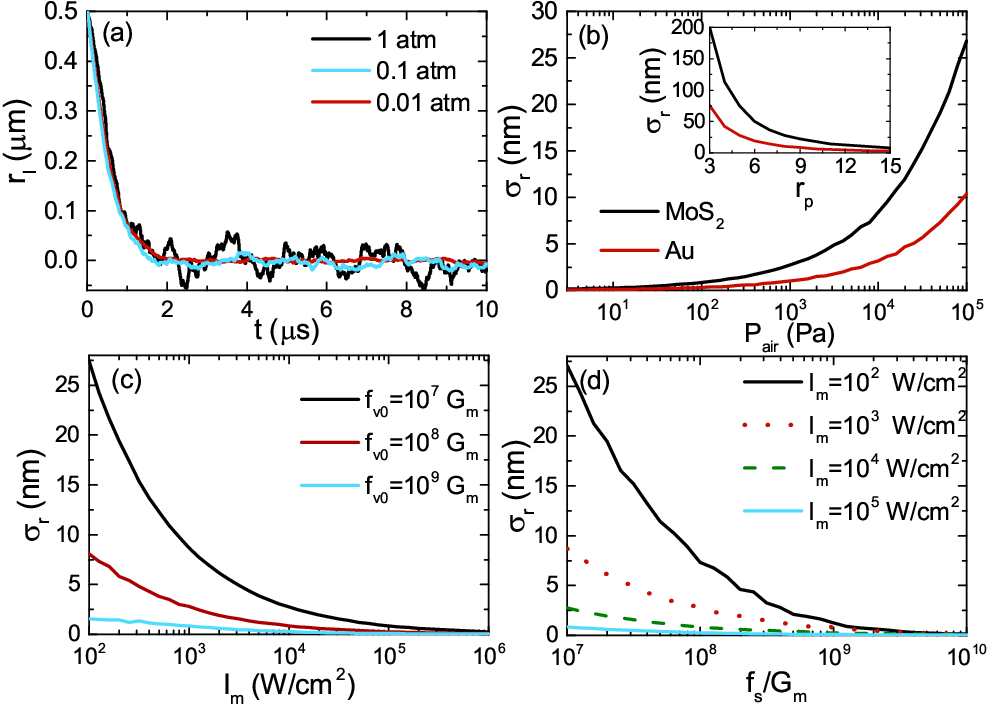}
\end{center}
\caption{(a) Nanoparticle's dynamics under different atmospheric pressures for MoS$_2$ nanoparticles. (b) Mean square displacement of MoS$_2$ or Au nanoparticles' position with respect to changes in atmospheric pressure, with an inset showing the mean square displacement concerning nanoparticle radius. (c) Mean square displacement of MoS$_2$ nanoparticles' position as a function of maximum light intensity for various dynamic friction constants, $f_{v0}$. (d) Mean square displacement of MoS$_2$ nanoparticles' position concerning changes in maximum light intensity for different dynamic friction constants, $f_{v0}$.}
\label{fig5}
\end{figure}

Due to the nanoparticle's rapid movement under the influence of the gradient force before it approaches the equilibrium position (0-1 $\mu s$), the displacement due to molecular collisions is minimal during this phase. Significant irregular motion only occurs once the nanoparticle approaches the steady state. As a result, we positioned the nanoparticle directly at the center ($t=0$) to study the mean square displacement caused by molecular collisions. To reduce the impact of randomness, each data point was calculated 100 times. The mean square displacement of position with changing air pressure is shown in Fig. 5(b). As air pressure decreases, the mean square displacement of position also decreases. For MoS$_2$ nanoparticles, at an air pressure of $10^3$ Pa and a maximum light intensity of 100 W/cm$^2$, the mean square displacement is approximately 2.7 nm, indicating minimal influence from Brownian motion. When the air pressure is $10^2$ Pa, the mean square displacement is around 0.8 nm, making the effect of Brownian motion essentially negligible. For Au nanoparticles, at an air pressure of $10^3$ Pa, the mean square displacement is approximately 1 nm, indicating minimal influence from Brownian motion. Like in single collisions, when nanoparticle radius increases, the nanoparticle's mass becomes larger, and the speed acquired during collisions becomes smaller. Additionally, the nanoparticle experiences a larger gradient force from the optical field. These factors combined result in a smaller mean square displacement of position. However, at 1 atmosphere, even with a radius of 15 nm, the mean square displacement for MoS$_2$ nanoparticles reaches 8.2 nm. At this point, the influence of Brownian motion cannot be completely ignored (insert of Fig. 5(b)).

Under 1 atmosphere of pressure, the influence of light intensity and dynamic friction constant ($f_{v0}$) is depicted in Figs. 5(c) and (d). Similar to the results from single collisions, higher light intensity provides a stronger gradient force, while a larger dynamic friction constant $f_{v0}$ enables faster dissipation of the kinetic energy transferred to the nanoparticle through air molecule collisions, thereby reducing the mean square displacement of position. When the dynamic friction constant $f_{v0}$ is $10^9 G_m$ ($\sim$26 pN), the light intensity in the range of $10^2-10^6$ W/cm$^2$ results in very small mean square displacements of position (less than 1.55 nm). When the maximum light intensity is $10^4$ or $10^5$ W/cm$^2$ and the dynamic friction constant $f_{v0}$ is in the range of $10^7-10^{10} G_m$, the mean square displacements of position remain very small (less than 2.8 nm). Furthermore, as seen in the results from Fig. 2(c), when the dynamic friction constant  $f_{v0}$ and light intensity are both relatively strong, it is more favorable for reducing the time it takes for the nanoparticle to reach the equilibrium position. Therefore, appropriately increasing the dynamic friction constant (e.g., $10^8G_m$) and light intensity (e.g., $10^4$ W/cm$^2$) is beneficial for overcoming the adverse effects of Brownian motion, speeding up processing, and facilitating practical applications for optically controlling superlubricated nanoparticles. It's worth noting that the required light intensity ($10^4$ W/cm$^2$) here is very low compared to the high light intensities needed in other optical force manipulation processes. For example, to overcome the strong interactions between traditional surfaces and micro/nanoparticles, a high peak laser power of up to $10^{11}$ W/cm$^2$ is required through a femtosecond laser system\cite{Shakhov2015}.

Finally, let's discuss how to utilize controlled nanoparticles for building surface nanostructures and the feasibility of such experiments. After manipulating nanoparticles to specific positions using optical trapping, it is possible to employ uniform beams of shorter wavelengths to heat the nanoparticles, causing them to lose their superlubricity and thus become immobilized in desired locations. This immobilization step is crucial for constructing specific lattice structures. Subsequently, the enhanced optical absorption of the photoresist near the nanoparticles, induced by nanoparticle surface plasmon resonance or scattering, can be utilized for etching corresponding microstructures. Alternatively, nanoparticles can serve as nucleation centers for the growth of nanostructures, achieved through techniques like molecular beam epitaxy (MBE).

Moreover, while this study focused on nanoparticles with extremely small radii of approximately 5 nm to showcase the super-resolution capabilities of optical manipulation on superlubricated surfaces, it's important to note that larger particles, ranging from tens of nanometers to micrometers in size, can also be effectively controlled on superlubricated surfaces. Manipulating larger particles becomes even easier due to the cubic dependence of the optical gradient force on particle radius. The key takeaway is that superlubricated surface optical manipulation has the potential to revolutionize the creation of surface nanostructures, offering unprecedented control and precision, even for larger nanoparticles.

In terms of experimental feasibility, the technology for manipulating micro- and nanoparticles using optical tweezers and other optical force methods is well-established. Simultaneously, the preparation and transfer techniques for superlubricated surfaces are also highly mature. This research effectively combines these two areas of expertise. Furthermore, the study's results indicate that the specific optical manipulation processes are robust and not highly sensitive to system parameters. For instance, when the dynamic friction constant, $f_{v0}$, varies between $10^5 G_m$ and
$10^9 G_m$, the system can be made to work efficiently by adjusting factors like optical intensity. What's particularly encouraging is that the required optical intensity is relatively low, with a maximum of only $10^4$ W/cm$^2$. This is significantly lower than the optical field strengths typically used in traditional optical tweezers. To put it into perspective, focusing 5 mW of laser power (equivalent to the brightness of a standard laser pointer) to the diffraction limit can already achieve an intensity of $10^5$ W/cm$^2$. Therefore, from an experimental standpoint, this research is entirely feasible.

 \section{Conclusions}

The research has revealed optical manipulation of nanoparticles on superlubricity surfaces. It was found that due to the absence of static friction on superlubricity surfaces, even nanoparticles with a radius as small as 5 nm can converge at the center of a diffracted light spot under the influence of a weak light field gradient, thus breaking the diffraction limit of light. Additionally, owing to the extremely low dynamic friction on superlubricity surfaces, the response time of nanoparticles is exceptionally fast, allowing particle manipulation to be accomplished within microseconds. Further reduction in the manipulation time can be achieved by increasing the dynamic friction constant (e.g., to $10^8 G_m$) and controlling the light intensity (e.g., $10^4$ W/cm$^2$). The effect of Brownian motion is significantly reduced as the manipulation takes place on a solid-solid interface and in a non-liquid environment. Adjusting the dynamic friction force and controlling the light intensity or reducing the air pressure can eliminate the impact of Brownian motion. Moreover, the system exhibits robustness to variations in parameters; for instance, when the dynamic friction constant ranges from $10^5 G_m$ to $10^9 G_m$, it can operate effectively by adjusting controllable factors such as light intensity. This excellent robustness is expected to facilitate practical implementations of optical super-resolution manipulation of nanoparticles. This breakthrough in optical manipulation on superlubricity surfaces holds significant potential for applications in fields such as photolithography, optical metamaterials, surface catalysis, and more.

\section*{Declaration of Competing Interest}
The authors declare that they have no known competing financial
interests or personal relationships that could have appeared to influence
the work reported in this paper.

\section*{Data availability}
Data will be made available on request.

\section*{Acknowledgments}
This work was supported by National Natural Science Foundation
of China (NSFC) (Grants No.62174040, No. 12174423), the 13th batch of outstanding young scientific and Technological
Talents Project in Guizhou Province (Grant no. [2021]5618).


\end{document}